\documentclass[aps,prb,showpacs]{revtex4}

\usepackage{amsmath}
\usepackage{color}
\usepackage{amssymb}

\usepackage{graphicx}

\begin{document}

\title{Half-quantum vortices in polar phase of superfluid $^3$He}

\author{V.P.Mineev}
\affiliation{Commissariat a l'Energie Atomique, INAC / SPSMS, 38054 Grenoble, France}

\begin{abstract}
The magnetic dipole-dipole interaction does not prevent existence of half-quantum vortices in the polar phase of superfluid $^3$He which can be stable in uniaxial anisotropic aerogel.  
Here we discus this exotic possibility. After developing a phenomenological theory of phase transition from the normal to polar superfluid state  and then to axipolar superfluid state
we calculate the NMR properties of such type superfluids at the rest or under rotation filled by an array either single quantum or half quantum vortices.
\end{abstract}
\pacs{ 67.30.he, 67.30.hm, 67.30.er}

\date{\today}
\maketitle

\section{ Introduction}
Unlike to superfluid $^4$He the  superfluid $^3$He-A supports the existence not only single quantum vortices but also the  vortices with half quantum of circulation \cite{Volovik1976}
and the superfluid flows with continuous distribution of vorticity.\cite{Mermin,Kopnin} The singular single quanta vortices as well as nonsingular vortices with 2 quanta of circulation 
 and also so called "vortex sheets" have been revealed in rotating $^3$He-A.\cite{Salomaa,Lounasmaa} However, the half quantum vortices in open geometry 
always possess an extra energy due to spin-orbit coupling preventing their formation. The discovery of vortices with half-quantum flux has been reported recently in mesoscopic samples of spin-triplet superconductor 
Sr$_2$RuO$_4$ similar to superfluid $^3$He-A. \cite{Jang}  Even earlier half-quantum vortices have been revealed in quite different ordered media -  an exciton-polariton condensate.\cite{Lagourdakis} In strontium ruthenate just the mesoscopic size of samples allows to avoid the problem of spin-orbital interaction whereas in exciton-polariton condensate we deal with completely different nature of the order parameter characterized by the light polarization and the phase of the condensate wave function.

The matrix of order parameter of superfluid $^3$He-A 
\begin{equation}
A^A_{\alpha i}=\Delta V_\alpha(\Delta^\prime_i+i\Delta^{\prime\prime}_i)/\sqrt{2}
\label{opa}
\end{equation}
is given by the product of its spin and orbital parts.\cite{Mineev}
 The unit spin vector ${\bf V}$ is situated in the plane perpendicular to the direction of spin up-up $|\uparrow\uparrow\rangle$ and down-down $|\downarrow\downarrow\rangle$ spins of the Cooper pairs. The vectorial product ${\bf \Delta}^\prime\times{\bf \Delta}^{\prime\prime}={\bf l}$
of orthogonal unit vectors  ${\bf \Delta}^\prime,{\bf \Delta}^{\prime\prime}$ determines the direction of the Cooper pairs orbital momentum  ${\bf l}$.

The half-quantum vortices are admissible because
a change of sign of orbital part of the order parameter acquired  over  any closed path in the liquid 
corresponding to half quantum vortex can be compensated by the change of sign of the spin part of the order parameter, 
so that the whole order parameter will be single-valued. 
These vortices in the superflow field are simultaneously disclinations in the magnetic anisotropy field ${\bf V}$ with half -integer Frank index, analogous to the disclinations in nematic liquid crystals.

More visual picture of half-quantum vortices can be given assuming that all vectors   change their directions living in $(x,y)$ plane: 
\begin{equation}
V_\alpha=\hat x_\alpha\cos\varphi/2-\hat y_\alpha\sin\varphi/2, ~~~\Delta^\prime_i+i \Delta^{\prime\prime}_i=(\hat x_i+i\hat y_i)e^{i\varphi/2}.
\label{hq}
\end{equation}
 Here $\varphi$ is the azimutal angle in $(x,y)$ plane perpendicular to the vortex axis along $\hat z$
 direction.
 The superfluid velocity is given by phase gradient  of the orbital part of the order parameter ${\bf v}_s=\frac{\hbar}{2m_3}\nabla\frac{ \varphi}{2}$. So, velocity circulation over a closed path 
$\gamma$ around $\hat z$ axis $\oint_\gamma{\bf v}_sd{\bf r}=\Gamma/2$ is twice smaller than it is for single quantum vortex with the order parameter distribution
\begin{equation}
V_\alpha=\hat z_\alpha,~~~\Delta^\prime_i+i \Delta^{\prime\prime}_i=(\hat x_i+i\hat y_i)e^{i\varphi}.
\label{sq}
\end{equation}
Here $\Gamma=\frac{h}{2m_3}=0,66\cdot 10^{-3} \frac{cm^2}{sec}$ is the circulation quantum.

Rotating vessel with a superfluid is filled by an array of parallel quantized vortices with density $n_v=2\Omega/\Gamma$ such that local average superfluid velocity is equal to velocity of normal component rotating with angular velocity 
$\Omega$ of vessel. The distance between the vortex lines is $r_v=n_v^{-1/2}$. Typical experimentally accessible speed of rotation is of the order several radians per second.  For instance
at $\Omega\approx 3~rad/sec$ the distance between the vortices is $\approx 10^{-2}cm.$

Energy of nonhomogeneous distribution of the order parameter is
\begin{equation}
{\cal F}_\nabla=\int d^3{\bf r}\left (K_1 \frac{\partial A_{\alpha i}}{\partial x_j}\frac{\partial A^\star_{\alpha i}}{\partial x_j}
+K_2 \frac{\partial A_{\alpha i}}{\partial x_j}\frac{\partial A^\star_{\alpha j}}{\partial x_i}
+K_3 \frac{\partial A_{\alpha i}}{\partial x_i}\frac{\partial A^\star_{\alpha j}}{\partial x_j}\right )
\end{equation}
Hence, the equilibrium energy density of rotating superfluid with angular velocity $\Omega$ filled by single quantum vortices Eq. (\ref{sq}) with density $n_v$ in logarithmic approximation is
given by the product of density of vortices  on the energy of one vortex
\begin{equation}
F_{n=1}=n_vf_{n=1},~~~~~f_{n=1}=\pi|\Delta|^2(2K_1+K_2+K_3)\ln\frac{r_v}{\xi}.
\end{equation}
Here $\xi$ is the coherence length.

The energy density of rotating superfluid with angular velocity $\Omega$ filled in equilibrium twice more half quantum vortices Eq. (\ref{hq})   does not have similar form. 
The reason for this is  the 
spin - orbital interaction caused by magnetic dipole interaction of helium nucleus. In a superfluid phase with triplet pairing  the density of SO coupling energy is \cite{Mineev}
\begin{equation}
F_{so}=\frac{g_D}{5|\Delta|^2}\left ( A_{\alpha\alpha}A_{\beta\beta}^\star+A_{\alpha i}A^\star_{i\alpha}-\frac{2}{3}A_{\alpha i}A^\star_{\alpha i} \right ),
\label{so}
\end{equation}
that  in case of A-phase with order parameter (\ref{opa})  is
\begin{equation}
F_{so}^A=\frac{g_D^A}{5}\left (
\frac{1}{3}-({\bf V}{\bf l})^2 \right ),
\end{equation}
where $g_D^A\propto |\Delta|^2$ is the constant of dipole-dipole interaction  proportional to square modulus of A-phase order parameter.
Hence, at distances larger than dipole length $\xi_d\sim 10^{-3}cm$ from the vortex axis the spin-orbital coupling suppress the inhomogeneity in the spin part of the order parameter distribution: vector $\bf V$ tends to be parallel or antiparallel to the direction of the Cooper pairs orbital momentum.  At these distances a disclination transforms in the domain wall (a planar soliton) possessing
energy proportional to its surface area.\cite{Volovik1976,Volovik1977,Mineev98}  The energy of rotating superfluid with angular velocity $\Omega$ filled by half quantum vortices Eq. (\ref{sq}) with density $2n_v$  have the following form
\begin{equation}
F_{n=1/2}=2n_vf_{n=1/2}=n_v\left [\pi |\Delta|^2(2K_1+K_2+K_3)\ln\frac{\xi_d}{\xi}+{\cal O}\left (|\Delta|^2K_{123}\frac{r_v}{\xi_d}  \right ) \right ],
\end{equation}
where
\begin{equation}
f_{n=1/2}=\frac{\pi}{2}|\Delta|^2(2K_1+K_2+K_3)\ln\frac{\xi_d}{\xi}+{\cal O}\left (|\Delta|^2K_{123}\frac{r_v}{\xi_d}  \right ).
\end{equation}
So, due to the energy of the domain walls connecting disclinations in vector ${\bf V}$ field an array  of  $2n_v$ half quantum vortices  is always nonprofitable in comparison with an array of $n_v$ single quantum vortices
\begin{equation}
F_{n=1/2}~>~
F_{n=1}.
\label{ineq}
\end{equation}

The neutralization of the dipole energy can be reached in the parallel plate geometry where $^3$He-A fills the space between the parallel plates with distance smaller than dipole length under  magnetic field $H>>25$ G applied along normal to the plates.\cite{Volovik1985,SalomaaPRL,Salomaa}
This case the half quantum vortices can energetically compete with single quantum  vortices. However, 
even in this case the  rotation of a "parallel plate" vessel with $^3$He-A will create lattice of half quantum vortices which at the same time presents  two-dimensional plasma of 
$\pm 1/2$ disclinations in the spin part of the order parameter with fulfilled condition of the "electroneutrality".\cite{Volovik1985,SalomaaPRL,Salomaa} The half quantum vortices in superfluid $^3$He-A till now have not been revealed.

 Fortunately the same magnetic dipole-dipole interaction does not prevent the existence of half-quantum vortices in the polar phase of superfluid $^3$He  probably realized \cite{Dmitriev12} in  peculiar porous media 
"nematically ordered" aerogel. We discuss the half quantum vortices in the polar state in the next chapter. Then a phenomenological description of the phase transition to the superfluid  polar state in porous media with uniaxial anisotropy will be presented.  Finally we shall derive the NMR  properties of rotating  polar state with an array of  half quantum vortices.
Some preliminary remarks on this subject have been published.\cite{Mineev2013}

\section{Half quantum vortices in polar state}

The polar state is another superfluid equal spin pairing state with an order parameter being product of spin  ${\bf V}$ and orbital ${\bf p}$ real unit vectors
\begin{equation}
A^{pol}_{\alpha i}=\Delta V_\alpha p_ie^{i\phi}.
\end{equation}
The order parameter distribution for a half quantum vortex is
\begin{equation}
\phi=\varphi/2,~~~~~~~{\bf p}=const,~~~~~~~~{\bf V}=\hat x\cos\frac{\varphi}{2}-\hat y\sin\frac{\varphi}{2},
\end{equation}
where $\varphi$ is azimuthal angle in $(x,y)$ plane perpendicular to the vortex axis along $\hat z$
 direction.

The spin-orbital energy Eq.(\ref{so}) in the polar state is
\begin{equation}
F_{so}=
\frac{2}{5}g^{pol}_D\left (({\bf V}{\bf p})^2-\frac{1}{3} \right ),
\end{equation}
where $g_D^{pol}\propto |\Delta|^2$ is the constant of dipole-dipole interaction  proportional to square modulus of polar state order parameter.
Thus at ${\bf V} \perp{\bf p}$ the spin-orbit interaction  does not play a role  and do not prevent formation of the half quantum vortices.
There is no need to work in parallel plates geometry and instead of inequality (\ref{ineq}) the energy densities of array of $n_v$ single quantum vortices
and array of $2n_v$ half quantum vortices are  proved to be equal in logarithmic approximation:
\begin{equation}
F_{n=1/2}=
F_{n=1}.
\label{eq}
\end{equation}

So, now there is no problem with spin-orbit coupling but where to get a liquid helium in superfluid polar state?

\section{Polar state in uniaxial anisotropic aerogel}

Let us consider an anisotropic aerogel with anisotropy axis 
along $\hat z$ filled by liquid $^3$He. By means the quasiparticle mean free path parallel 
$l_\parallel$ and perpendicular $l_\perp$ to anisotropy axis one can distinguish between the "easy axis" $l_\parallel>l_\perp$ and "easy plane" $l_\parallel<l_\perp$ anisotropy.
It is known that  anisotropy lifts the degeneracy between  the different  superfluid phases with p-wave pairing. \cite{Aoyama}
Namely, an easy plane anisotropy  stimulate the phase transition from normal state to the superfluid A-phase with the Cooper pair angular momentum $\vec l$ having preferable orientation along the anisotropy axis. On the contrary an easy axis anisotropy makes preferable phase transition from normal state to the superfluid polar state. At further  temperature decreasing  the latter is transformed to the axipolar state which at further decrease of temperature is transformed to the almost pure A-phase with vector $\vec l$ randomly distributed in the basal plane, that is so called Larkin-Imry-Ma (LIM) state.

The simplest phenomenological description of this phase transition sequence can be obtained making use the following Landau free energy functional
\begin{eqnarray}
&F_{cond}=\alpha A_{\alpha i}^\star A_{\alpha i}+\eta_{ij}A_{\alpha i}A^\star_{\alpha j}\nonumber\\
&+\beta_1|A_{\alpha i}A_{\alpha i}|^2+\beta_2A_{\alpha i}^\star A_{\alpha j}A_{\beta i}^\star A_{\beta j}+
\beta_3A_{\alpha i}^\star A_{\beta i}A_{\alpha j}^\star A_{\beta j}+\beta_4(A_{\alpha i}^\star A_{\alpha i})^2+\beta_5A_{\alpha i}^\star A_{\beta i}A_{\beta j}^\star A_{\alpha j}
\label{GL}
\end{eqnarray}
Here, $$\alpha=\alpha_0(T-T_{c}), $$ $T_c=T_c({P})$ is the transition temperature in superfluid state suppressed   in respect of transition temperature in the bulk liquid  $T_{c0}({P})$ due to isotropic part of quasiparticles scattering on aerogel strands. 

Media  uniaxial anisotropy with anisotropy axis 
along $\hat z$  is described phenomenlogically by  the traceless tensor
\begin{equation}
 \eta_{ij}=\eta\left (\begin{array}{ccc}1&0&0\\
 0&1&0\\
 0&0&-2\end{array}\right )
 \label{eta}
\end{equation}
which modifies  the second order terms in free energy.
The case of "easy axis" anisotropy corresponds to the positive coefficient $\eta>0$.
The quasiparticle scattering in an aerogel certainly changes the coefficients $\beta_i$ in respect to its value in the bulk liquid.
Moreover, the anisotropy of scattering leads to appearance of the new fourth order terms like 
\begin{equation} 
\beta^\prime_2A_{\alpha z}^\star A_{\alpha j}A_{\beta z}^\star A_{\beta j~~}+~~c.c. 
\label{anis}
\end{equation}
In the following for simplicity we shall ignore the additional anisotropic  fourth order terms (see, however, the remark at the end of this section). At final stage we shall use  the  values of $\beta_i$ coefficients established in weak coupling theory including strong coupling corrections found for the bulk liquid $^3$He stabilizing A-phase in respect to B-phase in high pressure-high temperature region of superfluid state phase diagram.

We shall work with the order parameter for the axipolar phase 
\begin{equation}
A_{\alpha i}=V_\alpha \left[a\hat z_i+ib(\hat x_i\cos\phi({\bf r})+\hat y_i\sin\phi({\bf r}))\right],
\label{op}
\end{equation}
here angle $\phi({\bf r})$ determines the local direction of product ${\bf \Delta}^\prime\times{\bf \Delta}^{\prime\prime}={\bf l}$ caused by local anisotropy according   Larkin-Imry-Ma ideology.
Substituting Eq.(\ref{op}) in Eq. (\ref{GL}) we obtain
\begin{equation}
F_{cond}=(\alpha-2\eta)a^2+(\alpha+\eta)b^2+\beta_{12}(a^2-b^2)^2+\beta_{345}(a^2+b^2)^2,
\label{fe}
\end{equation}
where
$$
\beta_{12}=\beta_1+\beta_2,~~~~~\beta_{345}=\beta_3+\beta_4+\beta_5.
$$
Minimization of Eq. (\ref{fe}) in respect of $a$ and $b$ gives that below the critical temperature
\begin{equation}
T_{c1}=T_c+2\frac{\eta}{\alpha_0}
\label{1}
\end{equation}
liquid passes from the normal state to the superfluid polar state with the order parameter amplitudes
\begin{equation}
a^2=a^2_0=-\frac{\alpha_0(T-T_{c1})}{2\beta_{12345}},~~~~b=0,
\label{polar}
\end{equation}
where 
$$\beta_{12345}=\beta_1+\beta_2+\beta_3+\beta_4+\beta_5.$$

Then below the critical temperature
\begin{equation}
T_{c2}=T_c-\frac{\eta}{\alpha_0}\frac{3\beta_{345}-\beta_{12}}{2\beta_{12}}
\label{c2}
\end{equation}
there is the second phase transition of the second order to the axipolar state with following amplitudes of the order parameter
\begin{equation}
a=a_0+\delta a,~~~~ \delta a=-\frac{\beta_{345}-\beta_{12}}{2\beta_{12345}}\frac{b^2}{a_0},~~~~b^2=-\frac{\alpha_0(T-T_{c2})}{4\beta_{345}}
\end{equation}
To estimate $T_{c2}$ let us use the values of  $\beta_i$ coefficients obtained in the weak coupling theory:
\begin{equation}
\beta_1=-0.5\beta,~~\beta_2=\beta,~~\beta_3=-\beta,~~\beta_4=\beta,~~\beta_5=\beta.
\label{sc}
\end{equation}
Here $\beta$ is a coefficient includes renormalization due to quasiparticle scattering in pure isotropic approximation. \cite{Thuneberg}
Its particular value plays no role because it drops out from the expression (\ref{c2}).
Thus, we obtain
\begin{equation}
T_{c2}=T_c-\frac{5}{2}\frac{\eta}{\alpha_0}.
\label{2}
\end{equation}
The strong coupling corrections lead to small modification of this formula, hence, Eqs. (\ref{1}) and (\ref{2}) show that  the critical temperature of normal - superfluid polar state transition is larger than the critical temperature in isotropic aerogel,
whereas the critical temperature of polar - axipolar state transition is smaller than the critical temperature in isotropic aerogel. Thus, we have the phase transition splitting
similar to that taking place in bulk A-phase under magnetic field splitting the superfluid phase transition to the sequence of two transitions in A$_1$ and then to A$_2 $ superfluid states.
Below $T_{c2}$ $a$ amplitude of the order parameter starts to decrease and $b$ amplitude is increased. At some temperature the $a$ and $b$ amplitudes will be almost equal each other
and superfluid ordering  acquires A-phase form.

When we work with free energy functional (\ref{GL}) at small anisotropy when $T_{c1}\approx T_{c2}$ we deal in fact with one phase transition from the normal to superfluid  with order parameter amplitudes $$a=b=\Delta_A/\sqrt{2}$$  and randomly distributed $\phi({\bf r})$ angle due to local anisotropy in Imry-Ma clusters. This case  we obtain
\begin{equation}
\Delta_A^2=-\frac{\alpha_0(T-T_{cA})}{2\beta_{345}},
\label{A}
\end{equation}
where
\begin{equation}
T_{cA}=T_c+\frac{\eta}{2\alpha_0,}.
\label{cA}
\end{equation}

One can distinguish with  what superfluid state we deal with by means of difference in NMR properties.

\section {Nuclear Magnetic Resonance properties}

The expression (\ref{so}) for the spin-orbital coupling in an  aerogel  with uniaxial anisotropy is modified to\cite{footnote}
\begin{equation}
F_{so}=C_1\left ( A_{\alpha\alpha}A_{\beta\beta}^\star+A_{\alpha i}A^\star_{i\alpha}\right )+C_2\left (2A_{zz}A_{\beta\beta}^\star+c.c.\right ) +const,
\label{soa}
\end{equation}
Here both coefficients $C_1$ and $C_2$ as well as the order parameter amplitudes depend on the quasiparticle scattering length in anisotropic aerogel.
Substituting Eq.(\ref{op}) in the above expression, performing averaging over the randomly distributed angle $\phi({\bf r})$ and  omitting the constant part  we obtain
\begin{equation}
F_{so}=[2(C_1+C_2)a^2-C_1b^2]\langle ({\bf V}(t)\hat z)^2\rangle.
\label{soan}
\end{equation}
Hence, in pure polar state (\ref{polar}) we obtain
\begin{equation}
F_{so}=2(C_1+C_2)\frac{\alpha_0(T_{c1}-T)}{2\beta_{12345}}\langle ({\bf V}(t)\hat z)^2\rangle.
\label{sopol}
\end{equation}

Here  the angular brackets mean time averaging over fast precessional motion of
${\bf V}(t)$ vector around direction of precessing magnetization with Larmor frequency $\omega_L$.\cite{Fomin,Gongadze}
Namely, if we have external magnetic field directed at the angle $\mu$ to the anisotropy axis $\hat z$ and the  precessing magnetization tilted on the angle $\beta$ to the field direction
\begin{equation}
{\bf V}(t)=R_x(\mu)R_z(-\omega_Lt)R_y(\beta)R_z(\omega_Lt+\varphi)\hat x,
\end{equation}
where $R_x(\mu)$ is the matrix of rotation  around $x$ axis on angle $\mu$ with similar definition of the other matrices.

Performing time averaging  we obtain  the angular dependence of the dipole energy
\begin{equation}
\langle ({\bf V}(t)\hat z)^2\rangle=\frac{1}{4}\sin^2\mu\left[(\cos\beta+1)^2\sin^2\varphi+(\cos\beta-1)^2/2\right]+\frac{1}{2}\cos^2\mu(1-\cos^2\beta).
\end{equation}

The frequency  shift  of transverse NMR from the Larmor value is given by  
\begin{equation}
2\omega\Delta\omega=-\frac{2\gamma^2}{\chi}\frac{\partial~\langle F_{so}\rangle|_{min}}{\partial \cos\beta}=-\Omega_{pol}^2\frac{\partial~\langle ({\bf V}(t)\hat z)^2\rangle|_{min}}{\partial \cos\beta},~~~~~~~~~~\Omega_{pol}^2=\frac{2\gamma^2}{\chi}2(C_1+C_2)\frac{\alpha_0(T_{c1}-T)}{2\beta_{12345}}.
\label{shift}
\end{equation}
Here $\gamma$ is helium-3 nucleus  gyromagnetic ratio, and $\chi$ is paramagnetic susceptibility.

 For space homogeneous distribution of ${\bf V}$ vector typical for the liquid at rest or in rotating liquid  filled by the single quantum vortices
 the minimum value of dipole energy $U_{D}|_{min}$ in respect of azimuthal angle $\varphi$  is reached at  $\varphi=0$.  This case the transverse NMR frequency shift is 
\begin{equation}
2\omega\Delta\omega=\Omega_{pol}^2\left (\cos\beta+\frac{1}{4}\sin^2\mu(1-5\cos\beta)\right ).
\end{equation}
On the other hand  for the space nonhomogeneous distribution of ${\bf V}$ vector typical for rotating liquid  filled by the half  quantum vortices
 one must average dipole energy $U_{D}|_{min}$ in respect of azimuthal angle $\varphi$.  This case the transverse NMR frequency shift  is 
\begin{equation}
2\omega\Delta\omega=\Omega_{pol}^2\left (\cos\beta-\frac{3}{2}\sin^2\mu\cos\beta\right ).
\end{equation}

Thus, different angular dependence of NMR frequency shift allows to distinguish single quantum and half quantum vortices arrays.

It is instructive to make formal comparison of the obtained frequency shift with the frequency shift  that occurs in case of direct transition to the Larkin-Imry-Ma A-phase.
This case
\begin{equation}
2\omega\Delta\omega=\frac{1}{2}\Omega_{A}^2\left (\cos\beta+\frac{1}{4}\sin^2\mu(1-5\cos\beta)\right ),~~~~~~~\Omega_{A}^2=\frac{2\gamma^2}{\chi}(C_1+2C_2)\frac{\alpha_0(T_{cA}-T)}{2\beta_{345}}.
\end{equation}
Here we used Eqs.(\ref{A}) and (\ref{soan}).  Then the slopes of linear temperature dependence of $\Omega_{pol}^2$ and $\Omega_{A}^2$ are related to each other as follows
\begin{equation}
\frac{d\Omega_{pol}^2}{d T}\left /\right.\frac{d\Omega_{A}^2}{d T}=\frac{2\beta_{345}}{\beta_{12345}}\frac{(C_1+C_2)}{C_1+2C_2}.
\end{equation}
Experimentally\cite{Dmitriev12} this ratio of slopes at small pressures is closed to 4/3 that corresponds to weak coupling value  of combination ${2\beta_{345}}/{\beta_{12345}}=4/3$ (see Eq.(\ref{sc})).
On the other hand the combination ${2\beta_{345}}/{\beta_{12345}}$ determined from other experiments \cite{Halperin} at different pressures takes values in interval from 1.25 to 1.39. \cite{footnote2}

Formula (\ref{soan}) also describes  the change of the temperature dependence of NMR frequency shift at phase transition from the pure polar state
to the axipolar state. Namely, at $T<T_{c2}$ we obtain
\begin{equation}
2\omega\Delta\omega=\Omega^2_{axipol}\left (\cos\beta+\frac{1}{4}\sin^2\mu(1-5\cos\beta)\right ),
\end{equation}
where  
\begin{equation}
\Omega^2_{axipol}=\frac{2\gamma^2}{\chi}2(C_1+C_2)\frac{\alpha_0(T_{c1}-T)}{2\beta_{12345}}\left [T_{c1}-T +r(T-T_{c2})\right  ]
\label{r}
\end{equation}
and
\begin{equation}
1-r=1-\frac{1}{4\beta_{345}}\left [2(\beta_{345}-\beta_{12})+\frac{C_1}{C_1+C_2}\beta_{12345}   \right ]
\end{equation}
is the ratio of the slopes of the temperature dependence of the phase shift below and above $T_{c2}$.
To estimate $r$  let us neglect by anisotropic part of spin-orbital energy, that is put $C_2=0$, and use the values of  $\beta_i$ coefficients in weak coupling approximation (\ref{sc}).
We obtain
\begin{equation}
r=\frac{3\beta_{345}-\beta_{12}}{4\beta_{345}}=\frac{5}{8}=0.625.
\end{equation}
Estimation using the data of the Ref.20 yields the r values  in interval from 0.6 to 0.64.\cite{footnote2}
Thus, the slope of $\Delta\omega$ temperature dependence below $T_{c2}$ begins to decrease.

\section{Conclusion}

The discovery of strongly uniaxial anisotropic aerogels opens new perspectives in physics of superfluid $^3$He.
First, it is an existence of pure polar state in temperature interval just below critical temperature of  normal-superfluid state  phase transition.\cite{Aoyama}
Second, it is a possibility to create an array of half quantum vortices in rotating vessel with the polar state filling this type of aerogel. 

It is shown that  half quantum vortices in the polar superfluid state  can be energetically stable due to absence of an extra energy of domain walls between them preventing their existence 
in superfluid $^3$He-A. 
In view of serious doubts to get reliable results   in frame of pure microscopic  theory, taking into account pressure dependent strong coupling corrections, anisotropy and randomness of quasiparticle scattering, we have chosen a pure
 phenomenological approach. It allows us to describe  the sequence of the phase transitions in the superfluid $^3$He filling uniaxial anisotropic aerogel and then to 
establish the transverse NMR properties for  fluid at rest as well as for the rotating superfluid, which were proved to be different in  case of single quantum and half quantum vortices filling the rotating vessel.

\section*{Acknowledgements}
I am indebted to V.V. Dmitriev for useful discussions.


\begin{thebibliography}{20} 

\bibitem{Volovik1976}G. E. Volovik and V. P. Mineev, Pis'ma Zh. Exp. Teor. Fiz. {\bf 24}, 605 (1976) [JETP Lett. {\bf 24}, 561 (1976)].

\bibitem{Mermin} N. D. Mermin, T.-L. Ho, Phys. Rev. Lett. {\bf 36}, 594 (1976).

\bibitem{Kopnin}G. E. Volovik and N. B. Kopnin,
Pis'ma Zh. Eksp. Teor. Fiz. {\bf 25}, 26 (1977) [JETP Lett. {\bf 25}, 22 (1977)]

\bibitem{Salomaa} M. M. Salomaa, G. E. Volovik, Rev. Mod. Phys. {\bf 59}, 533 (1987).

\bibitem{Lounasmaa} O.V.Lounasmaa and E. Thuneberg, PNAS {\bf 96}, 7760 (1999).

\bibitem{Jang} J. Jang, D. G. Ferguson, V. Vakaryuk, R. Budakian, S. B. Chung, P. M. Goldbart, Y.Maeno, Science, {\bf 331}, 186 (2011).

\bibitem{Lagourdakis}K. G. Lagourdakis, T. Ostatnicky, A. V. Kavokin, Y. G. Rubo, R. Andre, B. Deveaud-Pledran, Science {\bf 326}, 974 (2009).

\bibitem{Mineev}V. P. Mineev, Usp. Fiz. Nauk. {\bf 139}, 303 (1983) [Sov. Phys. Usp. {\bf 26}, 160 (1983)].

\bibitem{Volovik1977}G. E.Volovik and V. P. Mineev, Zh. Exp. Teor. Fiz. {\bf 72}, 2256 (1977) [Sov. Phys. JETP {\bf 45}, 1186 (1977)]. 

\bibitem{Mineev98}V. P. Mineev, {\it Topologically Stable Dedects and Solitons in Ordered Media"},  Harwood Academic Publishers, 1998.

\bibitem{Volovik1985}G. E. Volovik and M. M. Salomaa, 
Zh. Eksp. Teor. Fiz. {\bf 88}, 1656 (1985)[Sov. Phys. - JETP {\bf 61}, 986
(1985)].

\bibitem{SalomaaPRL} M. M. Salomaa M and Volovik G. E., Phys. Rev. Lett.,
{\bf 55}, 1184  (1985).

\bibitem{Dmitriev12} R. Sh. Askhadullin, V. V. Dmitriev, D. A. Krasnikhin, P. N. Martynov, A. A. Osipov, A. A. Senin, A. N. Yudin, PisÕma v ZhETF, {\bf 95}, 355 (2012); JETP Lett. {\bf 95}, 326
(2012). 

\bibitem{Mineev2013} V.P.Mineev, Low Temp. Phys. / Fizika Nizkikh Temperatur {\bf 39}, 1056 (2013).

\bibitem{Aoyama} K. Aoyama and R. Ikeda, Phys. Rev. B {\bf 73}, 060504¨ (2006).


\bibitem{Thuneberg}E. V. Thuneberg, S. K. Yip, M. Fogelstrom, and J. A. Sauls, Phys. Rev. Lett. {\bf 80}, 2861, (1998). 


\bibitem{footnote} In fact the second "anisotropy" term in  formula (\ref{soa}) is valid for for p-wave superfluid with an order parameter being the product of spin and orbital parts. In general  it is somewhat more compex.

\bibitem{Fomin} I.A.Fomin, Zh.Eksp.Teor.Fiz. {\bf 71}, 791 (1976)[ Sov. Phys. JETP {\bf 44}, 416 (1976)].

\bibitem{Gongadze} A.D.Gongadze, G.E.Gurgenishvili, and G.A.Kharadze, Zh. Eksp. Teor. Phys. {\bf 78}, 615 (1980); Sov. Phys. JETP {\bf 51}, 310 (1980).

\bibitem{Halperin} H.Choi, J. P. Davis, J. Pollanen, T. M. Haard, and W. P.Halperin, Phys, Rev. B {\bf 75}, 174503 (2007).

\bibitem{footnote2} V.V.Dmitriev, privite communication. Note, that to calculate the numerical values for different combinations of $\beta_i$ coefficients one has to take into account the difference in definitions of  $\beta_i$  in present paper and $\beta_i^H$ in Ref. 20. Namely, there is the following correspondence
$$
\beta_1=\beta_1^H,~\beta_2=\beta_3^H,~\beta_3=\beta_5^H,~\beta_4=\beta_2^H,~\beta_5=\beta_4^H.
$$







\end{thebibliography}
\end{document}